\documentclass[doublecol]{epl2}

\usepackage{graphicx}
\usepackage{amssymb}
\usepackage{amsmath}
\usepackage{amsfonts}
\usepackage{overpic}
\usepackage{color}
\usepackage{bm}

\DeclareMathAccent{\ring}{\mathalpha}{operators}{"17}
\providecommand{\st}[1]{_{\text{#1}}}

\def\onehalf{\frac{1}{2}}

\def\bra{\ensuremath{\langle}}
\def\ket{\ensuremath{\rangle}}

\def\const{\mathrm{const}}

\def\Rv{\bv{R}}

\def\b0{\bv{0}}

\def\ra{\rightarrow}

\def\msd{\bra \Delta R_\alpha(t)^2\ket}
\def\uRMS{\bra \delta u^2\ket^\onehalf}
\newcommand{\mods}{\ensuremath{\kappa_{\text{S}}}}
\newcommand{\modb}{\ensuremath{\kappa_{\text{B}}}}

\newcommand{\TBTT}{tumbling-to-tank-treading transition}

\newcommand{\bitem}{\begin{itemize}}
\newcommand{\eitem}{\end{itemize}}
\newcommand{\benum}{\begin{enumerate}}
\newcommand{\eenum}{\end{enumerate}}
\newcommand{\bblock}[1]{\begin{block}{#1}}
\newcommand{\eblock}{\end{block}}
\newcommand{\bmini}[1]{\begin{minipage}{#1}}
\newcommand{\emini}{\end{minipage}}
\newcommand{\btab}[1]{\begin{tabular}{#1}}
\newcommand{\etab}{\end{tabular}}
\newcommand{\btabn}[1]{\begin{tabular}{#1}}
\newcommand{\etabn}{\end{tabular}}
\newcommand{\beq}{\begin{equation}}
\newcommand{\eeq}{\end{equation}}
\newcommand{\beqn}{\begin{equation*}}
\newcommand{\eeqn}{\end{equation*}}
\newcommand{\bmult}{\begin{multline}}
\newcommand{\emult}{\end{multline}}
\newcommand{\bsplit}{\begin{split}}
\newcommand{\esplit}{\end{split}}

\newcommand{\bv}[1]{\mathbf{#1}}
\newcommand{\Ca}{\ensuremath{\text{Ca}}}

\title{Fluctuations and diffusion in sheared athermal suspensions of deformable particles}
\author{Markus Gross\inst{1,2,3}\thanks{Corresponding author: \email{gross@is.mpg.de}} \and Timm Kr\"uger\inst{4} \and Fathollah Varnik\inst{1,5}}
\shortauthor{M.\ Gross \etal}

\institute{                    
	\inst{1} Interdisciplinary Centre for Advanced Materials Simulation (ICAMS), Ruhr-Universit\"at Bochum, Universit\"atsstra{\ss}e 150, 44780 Bochum, Germany\\
	\inst{2} Max-Planck-Institut f\"{u}r Intelligente Systeme, Heisenbergstra{\ss}e 3, 70569 Stuttgart, Germany\\
	\inst{3} Institut f\"{u}r Theoretische Physik IV, Universit\"{a}t Stuttgart, Pfaffenwaldring 57, 70569 Stuttgart, Germany\\
	\inst{4} School of Engineering, The University of 
	Edinburgh, Edinburgh EH9 3JL, United Kingdom\\
	\inst{5} Max-Planck Institut f\"ur Eisenforschung, Max-Planck Stra{\ss}e~1, 40237 D\"usseldorf, Germany
}


\abstract{
We analyze fluctuations of particle displacements and stresses in a sheared athermal suspension of elastic capsules (red blood cells). 
Upon variation of the volume fraction from the dilute up to the highly concentrated regime, our numerical simulations reveal different characteristic power-law regimes of the fluctuation variances and relaxation times.
In the jammed phase and at high shear rates, anomalous scaling exponents are found that deviate from pure dimensional predictions.
The observed behavior is rationalized via kinetic arguments and a dissipation balance model that takes into account the local fluid flows between the particles.
Our findings support the view that the rheology of dense suspensions is essentially governed by the non-affine displacements.
}

\begin{document}

\maketitle

\section{Introduction}

Athermal suspensions of deformable particles, such as blood, occur manifold in nature and technology and a statistical description of these far-from-equilibrium systems is a challenging task. 
Characteristic for their behavior is their ability to `jam', that is, to acquire a solid-like character above a certain critical particle concentration, yet they remain amorphous in structure \cite{van_hecke_jamming_2010}. 
Any dynamics in an athermal suspension can only be induced by external means, such as shearing.
Under constant shear, athermal suspensions generally show diffusive behavior \cite{eckstein_self-diffusion_1977, leighton_shear-induced_1987} and the resulting mixing and segregation phenomena are crucial for the flow of blood \cite{higgins_statistical_2009, kumar_mechanism_2012}, drug delivery and the processing of pasty materials as divers as clay, food and cosmetic items \cite{van_der_sman_effective_2012}. 
While shear-induced diffusion is well understood for suspensions of rigid particles \cite{acrivos_longitudinal_1992, brady_microstructure_1997, breedveld_measurement_1998, drazer_deterministic_2002, sierou_shear-induced_2004, pine_chaos_2005},
soft-particle suspensions consisting of vesicles or capsules have so far been investigated almost exclusively in the \emph{dilute} regime, where hydrodynamic two-particle interactions dominate \cite{lac_hydrodynamic_2007,lac_pairwise_2008, le_hydrodynamic_2011, zhao_shear-induced_2011, tan_hydrodynamic_2012,kruger_interplay_2013}.
For a \emph{dense} suspension, however, collective effects and near-contact interactions are prevalent \cite{van_hecke_jamming_2010}.
Moreover, in the case of driven glassy and jammed systems, a range of different predictions and observations for the scaling behavior of the diffusivity with shear rate or volume fraction exists \cite{besseling_three-dimensional_2007, lemaitre_rate-dependent_2009, olsson_diffusion_2010, eisenmann_shear_2010, mobius_relaxation_2010, heussinger_superdiffusive_2010}. 

In the present work, we study shear-induced diffusion together with the underlying particle displacement and stress fluctuations in a \emph{dense} athermal suspension of aggregation-free red blood cells (RBCs) under wall-driven shear flow. 
We cover volume fractions $\phi$ between 12\% and 90\% and four orders of magnitude in reduced shear rate, which include not only the typical human body conditions ($\phi\sim 40-45\%$), but also regimes that are relevant for certain diseases -- such as thrombosis or polycythemia (large $\phi$) -- as well as for microfluidic processing of blood (low $\phi$, \cite{davis_deterministic_2006}). 
We identify distinct scaling behaviors in the dilute and in the jammed regime: in the dilute case, fluctuations typically scale `canonically' with shear rate, i.e., they follow the predictions from purely dimensional arguments. In contrast, `anomalous' scaling is observed in the jammed phase. 
The static and dynamic behavior of the displacement and stress fluctuations is quantitatively explained in terms of a dissipation balance, which links the local shear rate fluctuations to the injected power. 
Our analysis extends previous applications \cite{ono_velocity_2003, andreotti_shear_2012} of dissipation arguments to more realistic suspension models that involve explicit solvent hydrodynamics and particle elasticity. The predictions are expected to be applicable to a wide range of soft-particle suspensions.
Together with simple kinetic arguments, we arrive, for the first time, at a complete characterization of fluctuations and diffusion in a suspension of hydrodynamically interacting elastic particles.

\section{Simulations and rheology}
\begin{figure}[t]\centering
	\begin{overpic}[width=0.30\linewidth]{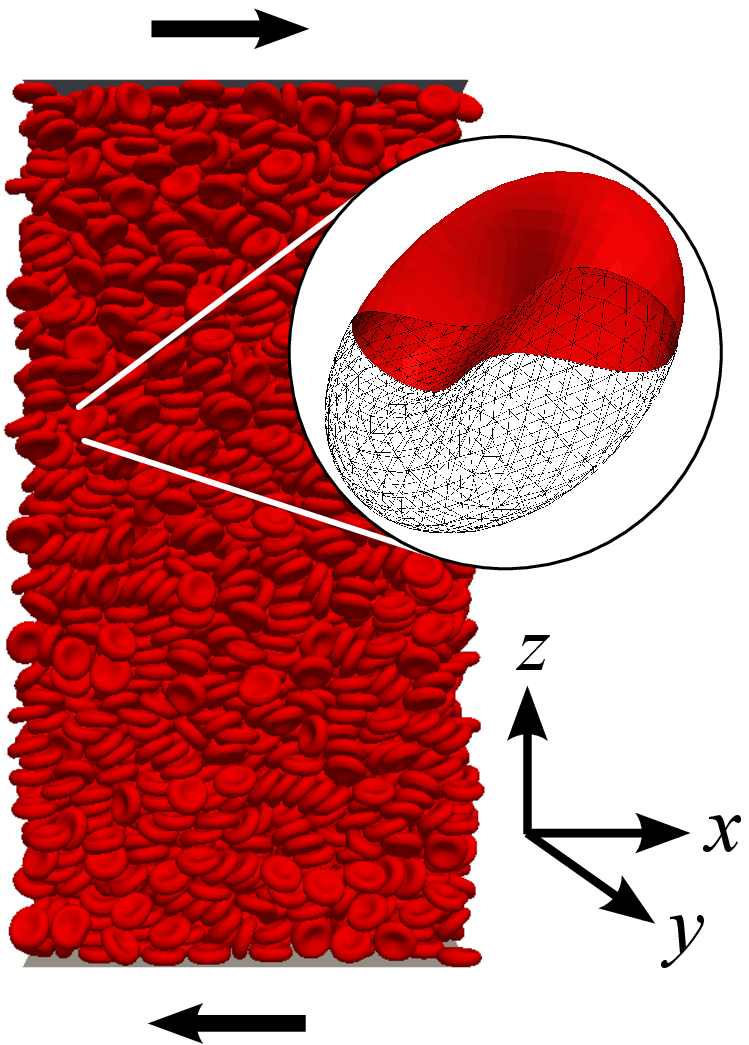}	\put(-2,-1){(a)}	\end{overpic}
	\begin{overpic}[width=0.66\linewidth]{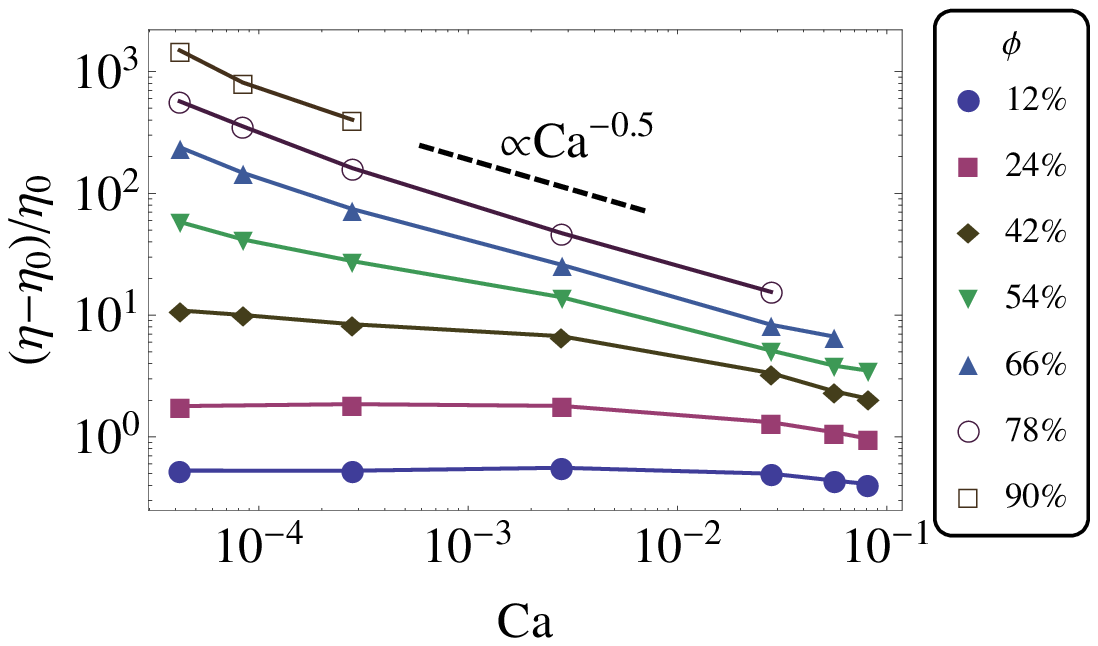}	\put(5,0){(b)}	\end{overpic}
	\caption{(a) Simulation setup including a sketch of the surface mesh of an RBC. The flow is bounded by walls in the $z$-direction and periodic in the $x$- and $y$-directions. Shear is applied in the $x$-direction (indicated by the black arrows). The system size is $L_x\times L_y\times L_z = 10\times 10\times 20$ particle diameters. The structure of the suspension remains amorphous over the whole parameter space studied. (b) Effective viscosity of the particle phase in dependence of the capillary number.}
	\label{fig_sim}
\end{figure}

\begin{table*}[t]\centering
	\begin{tabular}{c|ccccccc}
		\hline\hline
		bare capillary number (Ca)     & 0.081 & 0.056 & 0.028 & 0.0028 & $2.8\times 10^{-4}$ & $8.4\times 10^{-5}$ & $4.2\times 10^{-5}$ \\
		shear rate $\dot\gamma/10^{-4}$ & 1.6   & 1.1   & 0.56  & 0.056  & 0.19                 & 0.11 & 0.056 \\
		shear modulus $\mods$   & 0.003 & 0.003 & 0.003 & 0.003  & 0.1   & 0.2 & 0.2 \\
		\hline\hline
	\end{tabular}
	\caption{Shear rate and elastic shear modulus (in lattice units) corresponding to the different bare capillary numbers used in this work. The bending modulus, which describes the resistance of the capsule to bending forces, is taken as $\modb=\mods/5$. The size of the simulation box is $L_x\times L_y\times L_z = 180\times 180\times 360$ lattice units and contains between 1000 and 7700 RBCs, each having a large radius of $r=9$ lattice units. The interior of a capsule is filled with a fluid of the same viscosity as the surrounding. See \cite{kruger_efficient_2011, kruger_crossover_2013, gross_rheology_2014} for a detailed description of the underlying simulation model.}
	\label{tab_param}
\end{table*}

RBCs in a Newtonian solvent are simulated via a combined Finite-Element-Immersed-Boundary-Lattice-Boltzmann method, which has been thoroughly benchmarked \cite{kruger_efficient_2011, kruger_particle_2011} and whose details can be found in \cite{kruger_crossover_2013, gross_rheology_2014}.
The capsules posses a shear and bending elasticity and are approximately area- and volume-incompressible. 
The interior of a capsule is filled with a fluid of the same viscosity as the surrounding.
Particles interact only via hydrodynamics and short-range repulsive forces, the latter of which are essentially present to avoid direct particle contacts and have been previously shown to be negligible for the rheology \cite{gross_rheology_2014}.
The shear rate $\dot \gamma$ is imposed by the walls (see Fig.~\ref{fig_sim}a) and is expressed in terms of the dimensionless capillary number, 
\beq \Ca= \frac{\eta_0 \dot\gamma r}{\mods}\,,
\label{eq_Ca}\eeq
where $\eta_0$ is the bare solvent viscosity, $r$ is the large RBC radius and $\mods$ is the shear-elastic modulus of the capsule. 
The choice of Ca as the governing parameter is motivated by the fact that the shear elasticity provides the most dominant contribution to the suspension stress \cite{gross_rheology_2014}.
Detailed simulation parameters can be found in Table~\ref{tab_param}.
We remark that we do not observe long-time stable shear bands \cite{mandal_heterogeneous_2012} or indications of crystallization.

As a central result we will show -- via computer simulations which take explicit account of hydrodynamic interactions -- the validity of dissipation-based theoretical arguments which relate fluctuation variances and relaxation times to the effective suspension viscosity $\eta=\eta(\phi,\Ca)$. The latter quantity has been discussed in \cite{gross_rheology_2014} and, for convenience, is shown again in Fig.~\ref{fig_sim}b. 
One may directly distinguish two different regimes: a Newtonian regime ($\eta\sim \text{const.}$) for small volume fractions ($\phi$) and a shear-thinning regime ($\eta\propto \Ca^{-q}$ with $q\simeq 0.5$) for larger $\phi$. A detailed numerical analysis of the stress data \cite{gross_rheology_2014} also reveals the existence of a third, yield-stress, regime ($\eta= \sigma_y \Ca^{-1} + b\Ca^{-q}$, with $\sigma_y$ and $b$ being $\phi$-dependent parameters), caused by the elastic compression of the capsules above a jamming concentration of $\phi_{\text{c}} \simeq 0.66$. In Fig.~\ref{fig_sim}b, this behavior is reflected by the slight upward bending of the viscosity curves for large $\phi$ and low Ca.

\section{Instantaneous particle velocity fluctuations}

\begin{figure}[t]\centering
	(a)\includegraphics[width=0.73\linewidth]{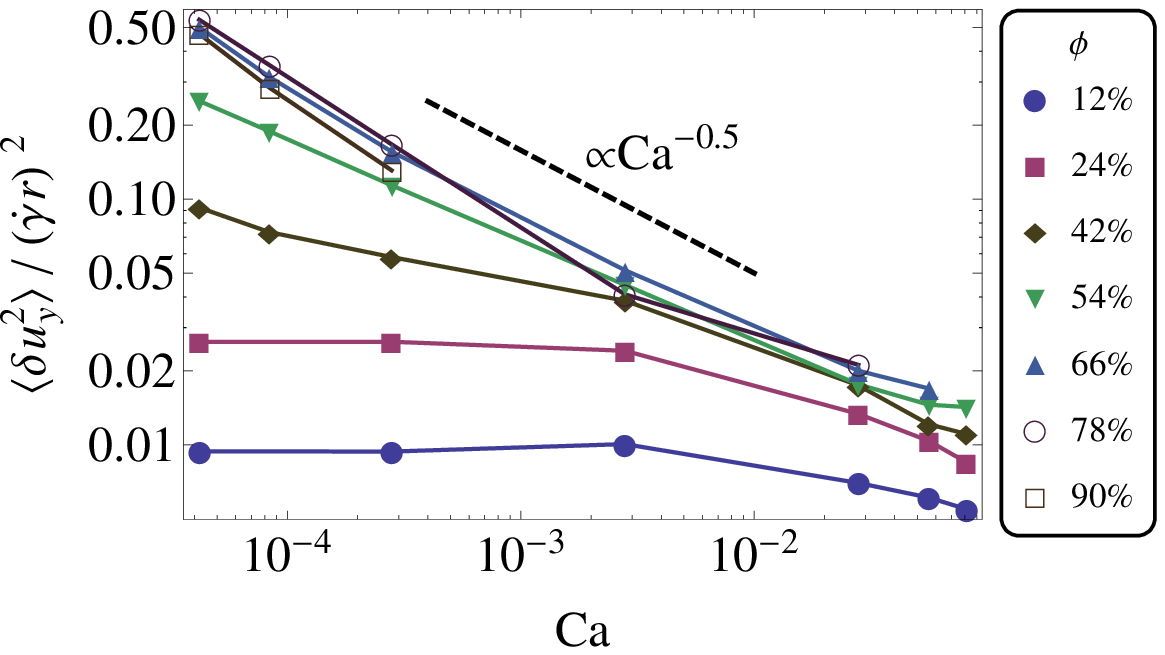}\quad
	(b)\includegraphics[width=0.73\linewidth]{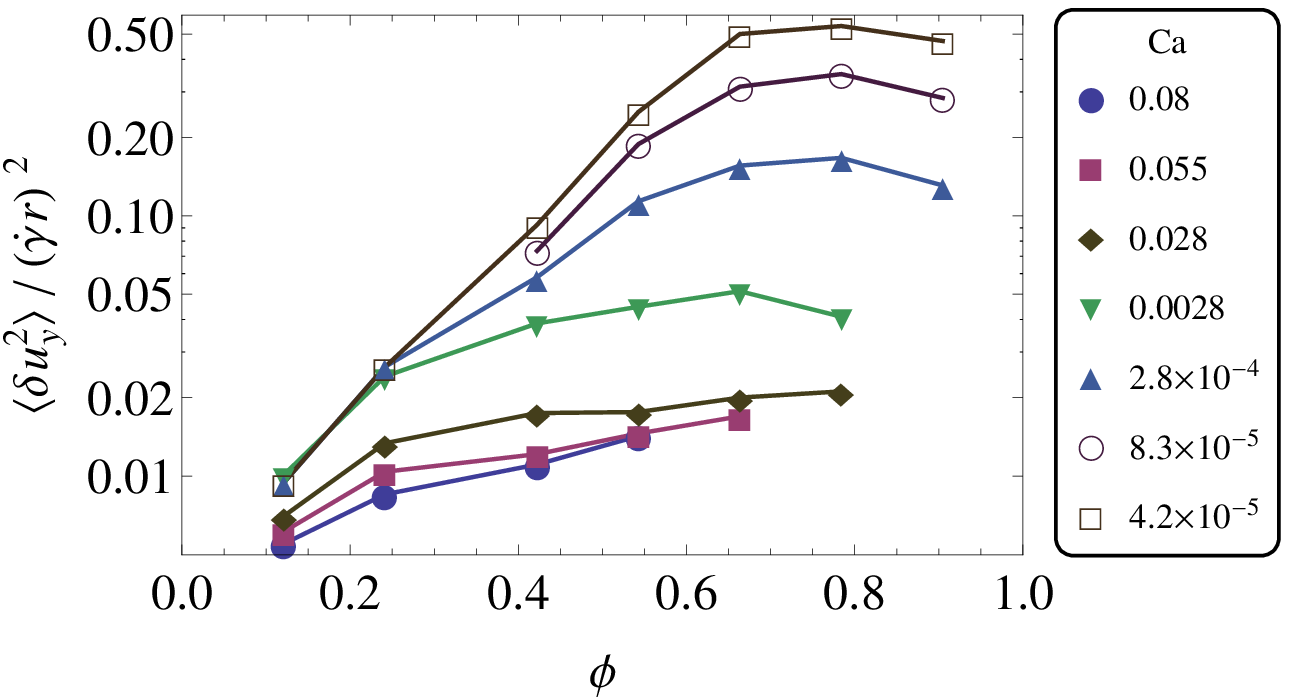}
	\caption{Variances of the instantaneous particle velocity fluctuations (displacements at vanishing strain) in the vorticity direction ($u_y$) in dependence of (a) the capillary number and (b) the volume fraction. The behavior is similar in the other two spatial directions $x$ and $z$ (not shown here).
	}
    \label{fig_velfluct}
\end{figure}

\begin{figure}[t]\centering
	\includegraphics[width=0.8\linewidth]{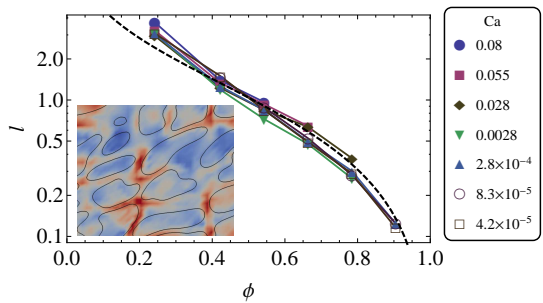}\quad
	\caption{Effective interparticle gap size $l$ as obtained from velocity fluctuations in our simulations via eq.~\eqref{eq_vel_scal} (symbols). The dashed curve represents the theoretical estimate of eq.~\eqref{eq_gap_size}, taking $r\simeq 2$ as a fit parameter. Inset: The magnitude of the viscous stress (red corresponding to large, blue to low values) as a measure for the local dissipation in the liquid. Large dissipation occurs in the narrow gaps between adjacent cells.}
    \label{fig_gap_scale}
\end{figure}

Fig.~\ref{fig_velfluct} shows the variance of the instantaneous particle velocity fluctuations  $\bra \delta u_y^2\ket$, scaled by the natural time and length scales, $1/\dot\gamma$ and $r$, in the vorticity direction ($y$). 
The probability distribution of $\delta u_y$ are found to have Gaussian cores with exponential tails that are enhanced for larger $\phi$ and lower Ca.
Here and in the following, averages are performed over time offsets and all particles in the bulk of the simulation box (i.e., excluding a region of two particle diameters adjacent to the walls). 
The overall scaling behavior is similar in the flow ($x$) and shear-gradient ($z$) directions and not separately shown.
From Fig.~\ref{fig_velfluct}a, we note a plateau at low Ca and $\phi$ and a characteristic power-law behavior at large Ca and $\phi$. 
As Fig.~\ref{fig_velfluct}b shows, the velocity fluctuations grow approximately exponentially with particle concentration up to $\phi \simeq 0.6-0.7$, beyond which they saturate or even slightly decrease. 
Notably, this cross-over volume fraction is close to the random-close-packing value of $\phi_c\simeq 0.66$ obtained for hard oblate ellipsoids of the same aspect ratio ($\sim 0.33$) as RBCs \cite{donev_improving_2004}. 
As the reduced instantaneous velocity fluctuation represents a length scale (the zero strain limit of a displacement), the non-monotonic volume fraction dependence is consistent with the idea that the limit $\phi\ra \phi_c$,  $\Ca\ra 0$ represents a critical point for an athermal suspension \cite{olsson_critical_2007, heussinger_superdiffusive_2010}.

Independently from the notion of a critical point, a quantitative and complete description of the particle velocity fluctuations can be achieved by noting that, for sufficiently large $\phi$, dissipation occurs predominantly in the fluid (occupying a volume fraction of $1-\phi$) between the particles (cf.\ \cite{mills_apparent_2009}). 
The typical shear rate in the gap of characteristic size $l$ between two neighboring particles can be expressed as $\dot\gamma_\alpha^*\simeq (\Delta \bar u_\alpha + \delta u_\alpha)/l$, where $\Delta \bar u_\alpha=\dot\gamma l\, \delta_{\alpha x}$ is the velocity difference due to the affine flow (only in $x$-direction) and $\delta u_\alpha$ is the velocity fluctuation. 
As the total power injected into the system is given by $\sigma \dot{\gamma}$ -- independently of the precise origin of $\sigma$ (i.e., viscous or elastic) -- dissipation balance requires that
$\eta\dot\gamma^2 = \eta_0\dot{\gamma}^2+(1-\phi)\eta_0 \sum_\alpha \bra \dot\gamma_\alpha^{*2}\ket\,,$
where first term arises from the affine motion of the background fluid and the sum runs over all three spatial directions.
To proceed, we assume that the average flow velocity and its fluctuations are uncorrelated ($\bra \Delta\bar u_x \delta u_x\ket = 0$) and obtain, as our first central result, a relation between the velocity fluctuations and the suspension viscosity
\beq \frac{\sum_\alpha \bra \delta u^2_\alpha\ket}{\dot\gamma^2} = \left[\frac{\eta-\eta_0}{(1-\phi)\eta_0} -1 \right]l^2 \,.
\label{eq_vel_scal}
\eeq
In the case of rigid spherical particles of radius $r$, the typical interparticle gap size can be estimated as 
\beq l \simeq 2r (\phi^{-1/3}-1 )\,.
\label{eq_gap_size}
\eeq
Note that we assumed here a maximal packing fraction of $\phi\st{max}=1$, which, for the present purpose, can be justified by the fact that short-range repulsive forces in our model prevent direct contacts between particles so that interstitial fluid regions are preserved.

As demonstrated in Fig.~\ref{fig_gap_scale}, for $\phi\gtrsim 0.2$, the estimate of eq.~\eqref{eq_gap_size} is consistent with our simulation data: taking $r\simeq 2$ (in lattice units, which is roughly 2/3 of the small radius of the RBC inertia ellipsoid) as a fit parameter -- which we deem admissible given the simplifications in our theory --, we observe an impressive agreement over almost the full parameter region studied. 
In particular, the independence of $l$ on capillary number is correctly reproduced. 
The breakdown of our scaling model at low $\phi$ is not surprising, as here $\eta\simeq \eta_0$ and dissipation becomes less localized.

\begin{figure*}[t]\centering
	\begin{overpic}[width=0.31\linewidth]{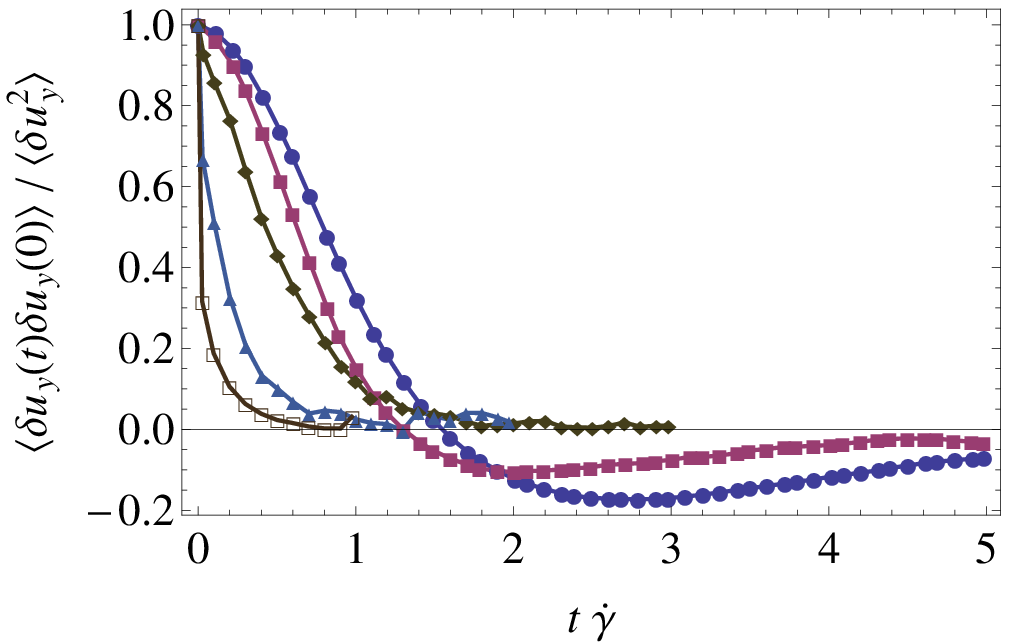}	\put(0,0){(a)}	\end{overpic}
	\begin{overpic}[width=0.34\linewidth]{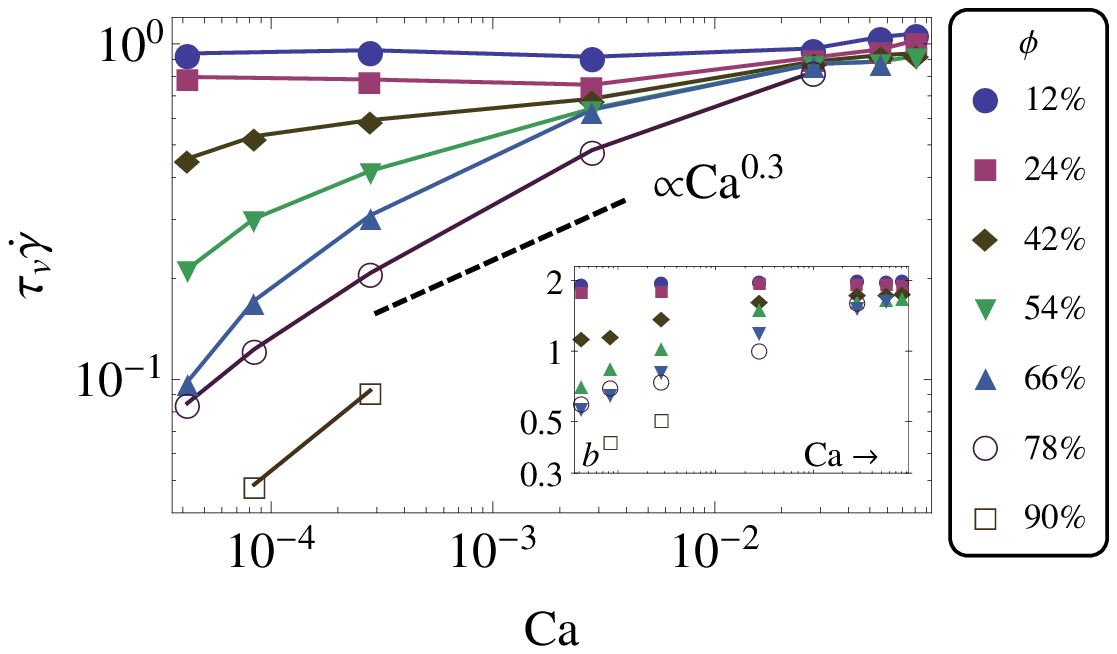}	\put(0,0){(b)}	\end{overpic}
	\begin{overpic}[width=0.335\linewidth]{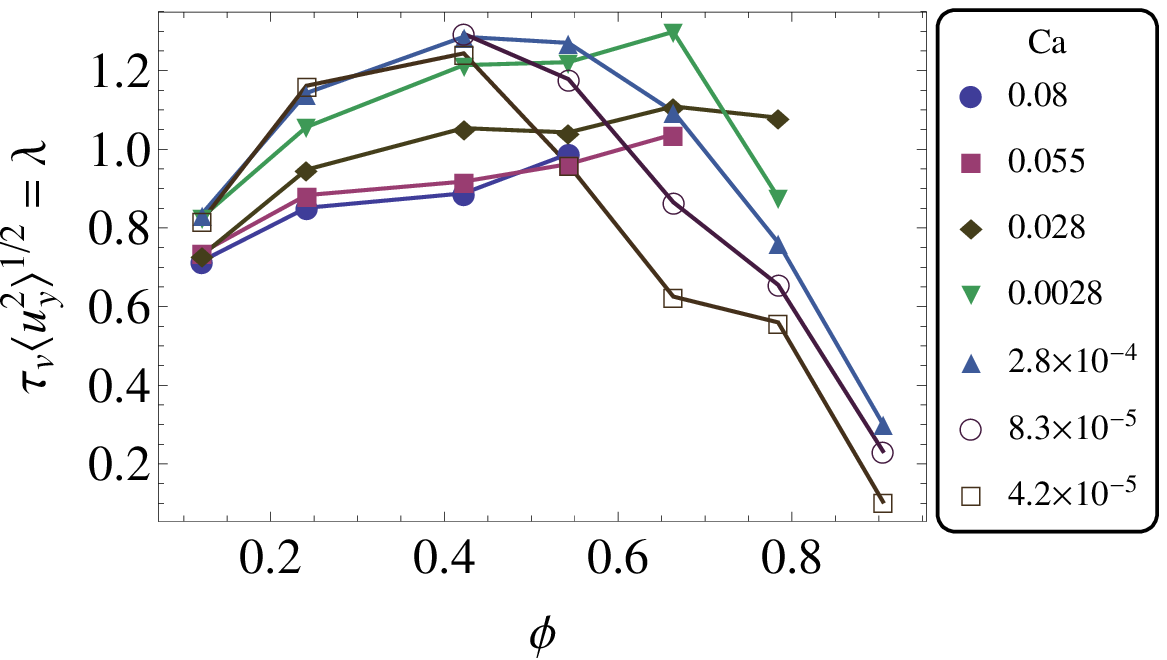}	\put(0,0){(c)}	\end{overpic}
	\caption{Velocity relaxation: (a) Typical behavior of the normalized VACF in our simulations [legend in panel (b) applies]. (b) Relaxation time $\tau_v$ and exponent $b$ (inset) obtained from fits of eq.~\eqref{eq_vac_form} to the VACFs in the vorticity direction. (c) Length scale $\lambda$ (`mean free path') defined by eq.~\eqref{eq_vel_tr}.}
	\label{fig_vel_relax}
\end{figure*}

In summary, relation \eqref{eq_vel_scal} indicates that the characteristic scaling behavior of the reduced velocity fluctuations observed in Fig.~\ref{fig_velfluct}a originates essentially from the effective viscosity (assuming $\eta_0\ll\eta$): indeed, we observe a Newtonian regime $\bra \delta u^2\ket/\dot\gamma^2 \propto \eta\sim \text{const.}$ for small $\phi$ and Ca and a shear-thinning regime $\bra \delta u^2\ket/\dot\gamma^2 \propto \eta\propto \Ca^{-q}$ with $q\simeq 0.5$ for large Ca. 
For smaller Ca, a yield stress regime is expected, where $\bra \delta u^2\ket/\dot\gamma^2 \propto \eta= \sigma_y \Ca^{-1} + b\Ca^{-q}$.
An unambiguous exhibition of this regime represents a formidable task for future work.

A similar connection between velocity fluctuations and rheology has also been previously noted for idealized jammed model systems in two dimensions \cite{ono_velocity_2003, andreotti_shear_2012}.
In these systems, however, particles were point-like and dissipation was explicitly implemented via a friction force that mimics the effect of the solvent. In contrast to this, the dynamics of the solvent is explicitly resolved by the lattice Boltzmann part of our simulation methodology. 
We furthermore infer from Fig.~\ref{fig_velfluct}a that relation \eqref{eq_vel_scal} holds independently for each velocity component.

\section{Velocity relaxation}

Collisions of two or more elastic particles in a sheared suspension break the reversibility of Stokes flow and give rise to diffusive behavior at long times \cite{loewenberg_collision_1997, lac_hydrodynamic_2007}.
The associated decorrelation of particle displacements can be quantified by the  velocity autocorrelation function (VACF), $C_v(t) = \bra \delta u_y(t)\, \delta u_y(0)\ket$, which is examplified in Fig.~\ref{fig_vel_relax}a.
Despite the negative long-time tail emerging at low volume fractions (which can be attributed to two-particle encounters \cite{drazer_deterministic_2002}), the VACF can be well fitted by a stretched exponential form, 
\beq C_v(t) = \bra \delta u^2\ket \exp[-(t/\tau_v)^b]\,,
\label{eq_vac_form}
\eeq
at small strains (i.e., $t\dot\gamma \lesssim 1$).
Fig.~\ref{fig_vel_relax}b shows the extracted relaxation time $\tau_v$.
The observed scaling behavior of $\tau_v$ can be explained by approximating
\beq \tau_v \simeq \lambda / \uRMS\,,
\label{eq_vel_tr}
\eeq
where $\uRMS$ is the typical particle velocity scale and $\lambda$ defines the characteristic distance a particle travels before its velocity changes significantly.
Assuming $\lambda$ depends only weakly on Ca, eq.~\eqref{eq_vel_tr} together with eq.~\eqref{eq_vel_scal} predicts $\tau_v \dot \gamma \sim \text{const.}$ in the Newtonian regime and $\tau_v \dot \gamma\propto \Ca^{q/2}$ in the shear-thinning regime, in approximate agreement with the simulation results (Fig.~\ref{fig_vel_relax}b).
As Fig.~\ref{fig_vel_relax}c shows, while its Ca-dependence is indeed weak, $\lambda$ varies by up to a factor of five over the range of volume fractions studied. 
The decrease of $\lambda$ with $\phi$ in the jammed phase is expected, as a higher packing density leads to a reduced mean-free path (kinetic theory would predict $\lambda\sim 1/\phi$).
The reason for the (albeit slight) increase of $\lambda$ with $\phi$ in the fluid regime, as well as for the behavior of the exponent $b$ (inset to Fig.~\ref{fig_vel_relax}b), is unclear at present.

\section{Mean-squared displacements and diffusivity}
\begin{figure*}[t]\centering
	\begin{overpic}[width=0.295\linewidth]{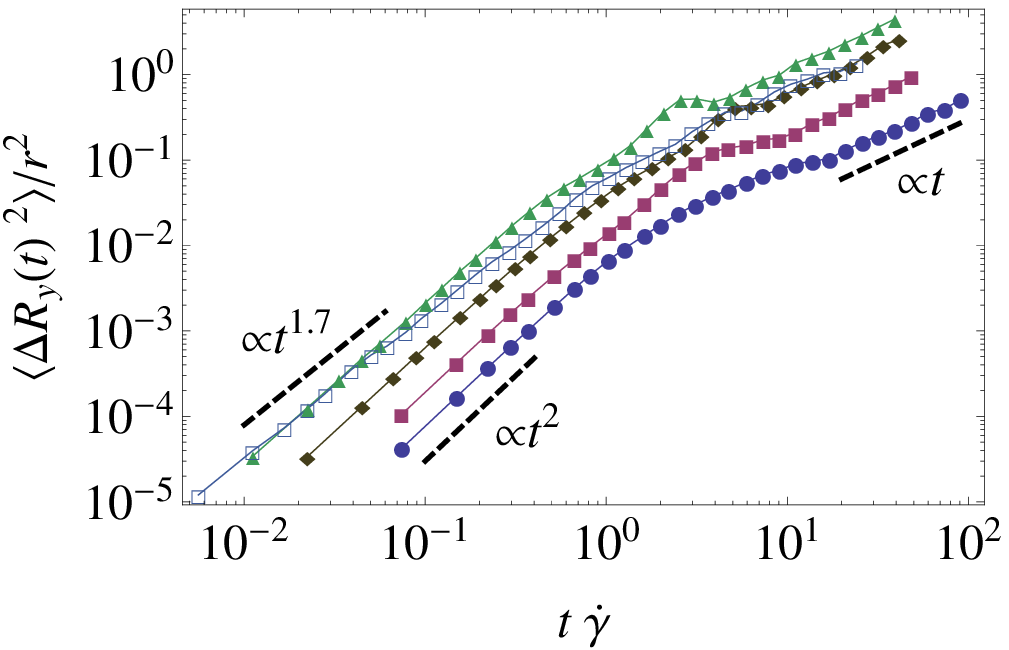}\put(0,0){(a)}\end{overpic}
	\begin{overpic}[width=0.335\linewidth]{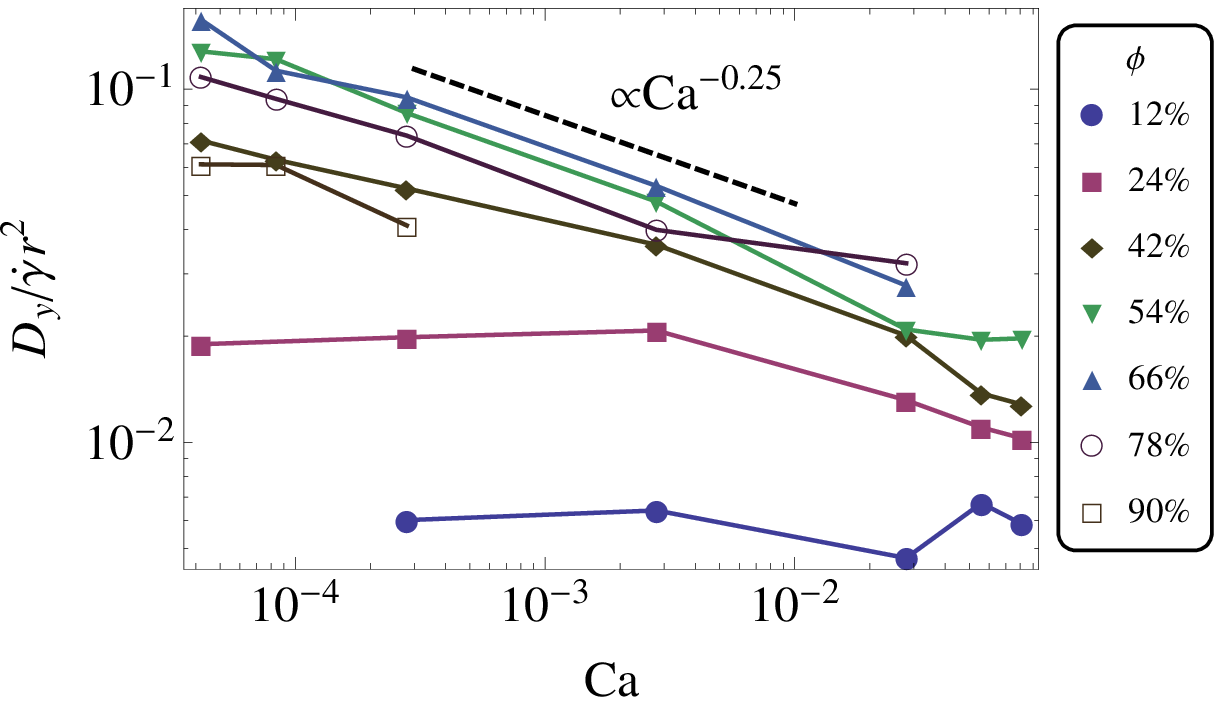}\put(0,0){(b)}\end{overpic}
	\begin{overpic}[width=0.345\linewidth]{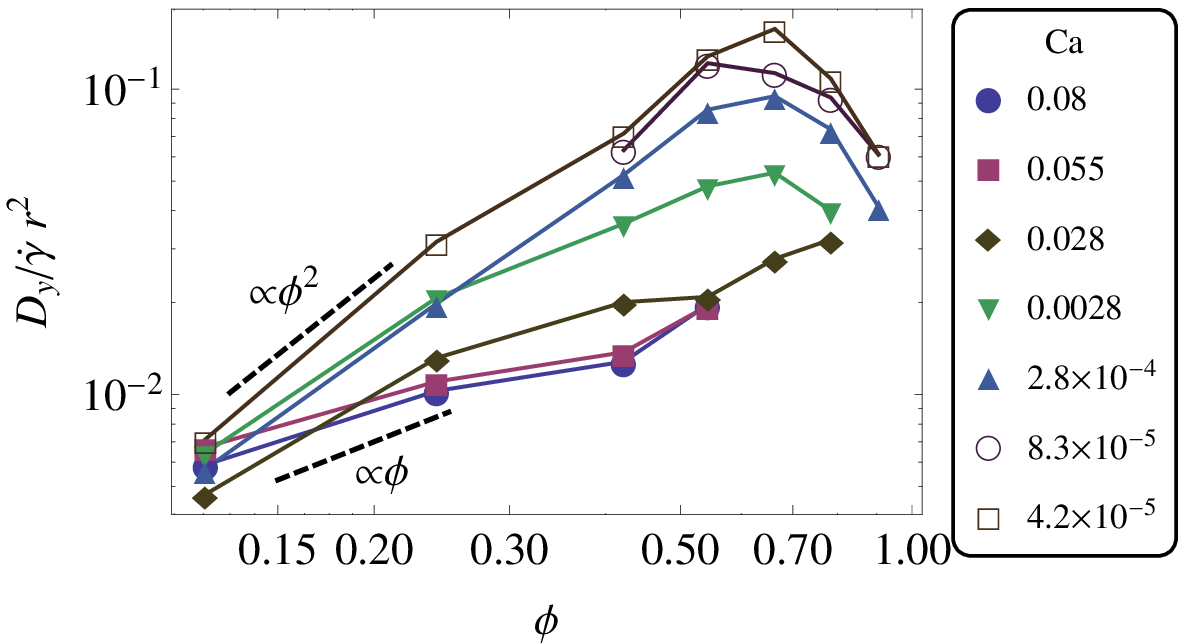}\put(0,0){(c)}\end{overpic}
	\caption{Mean-squared displacements and diffusivity: (a) Typical MSDs observed in our simulations [legend in panel (b) applies]. (b,c) Diffusion coefficient $D$ (in the vorticity direction), scaled by the intrinsic time and length scales, $1/\dot \gamma$ and $r$ in dependence of (b) capillary number and (c) volume fraction. (NB: For the data point at $\phi=24\%$ and the lowest Ca we have used simulation parameters $\dot{\gamma}=0.0056\times 10^{-4}$, $\mods=0.02$, deviating from Table~\ref{tab_param}.) }
    \label{fig_diff}
\end{figure*}
Fig.~\ref{fig_diff}a shows simulation results of typical mean-squared displacements (MSDs), $\bra \Delta R_\alpha(t)^2\ket = \bra [R_\alpha(t)-R_\alpha(0)]^2\ket$, with $\Rv$ being the center coordinate of a particle. 
Consistent with previous simulations of non-Brownian suspensions \cite{drazer_deterministic_2002, sierou_shear-induced_2004}, we observe a `ballistic' regime ($\msd\propto t^n$ with $n\simeq 2$) at small strains that crosses over into a diffusive regime ($\msd \propto t$) at large strains.
At the largest volume fraction studied ($\phi=0.9$), an exponent of $n\simeq 1.7$ is found in the ballistic regime, which has been associated with dynamical heterogeneities \cite{tanguy_plastic_2006, heussinger_superdiffusive_2010, mobius_relaxation_2010}.
The onset of a plateau connecting the ballistic and diffusive regimes of the MSD, concomitant to a negative long-time tail of the VAC, appears only at low volume fractions and can be attributed to two-particle encounters \cite{drazer_deterministic_2002}. 
We defer a more detailed discussion of these phenomena to a forthcoming work.  

The overall behavior of the MSDs can be understood from their relation to the VACF \cite{hansen_theory_2006}:
\beq \bra \Delta R(t)^2 \ket = 2 t \int_0^t\left(1-\frac{s}{t}\right)C_v(s)ds \,,
\label{eq_msd_vac}
\eeq
which predict a ballistic and diffusive regime,
\beq \bra \Delta R(t)^2 \ket \sim 
\begin{cases} t^2 C_v(0) \quad &(t\ra 0) \\ t \int_0^\infty C_v(s) ds \quad & (t\ra \infty)\,.
\end{cases}
\label{eq_msd_asymp}
\eeq
Importantly, these relations hold independently of the thermal or athermal nature of the system. In particular, the ballistic scaling of the MSD merely reflects the fact that particles move unperturbed at small times.
From this point of view, one might suspect that our observation of an exponent $n<2$ at early times is related to insufficient temporal resolution and that an exponent closer to 2 should emerge at sufficiently small times.

The diffusivity in the vorticity direction, as extracted from linear fits to the MSDs, is plotted in Figs.~\ref{fig_diff}b,c. The behavior in the shear-gradient direction is similar, although the relative magnitude of the diffusivity is larger (smaller) for low (high) $\phi$.
The observed scaling behavior can be rationalized from eq.~\eqref{eq_msd_asymp}, which yields, after making use of eqs.~\eqref{eq_vac_form} and \eqref{eq_vel_tr}:
\beq D=\int_0^\infty C_v(t) \mathrm{d}t = \bra \delta u^2\ket \tau_v \Gamma(1+1/b) \simeq \bra \delta u^2\ket^\onehalf \lambda\,.
\label{eq_D_scal}
\eeq
Here, $\Gamma$ is the Gamma-function, which we have neglected in the final result as it varies only weakly over the studied parameter ranges. 
Note that expression \eqref{eq_D_scal} applies also to a thermal equilibrium system if $\bra u^2\ket$ is taken as the kinetic temperature.
Eq.~\eqref{eq_D_scal} predicts  $D/\dot{\gamma}\sim \text{const.}$ in the Newtonian regime and $D/\dot{\gamma}\sim \Ca^{-q/2}$ in the shear-thinning regime, which is indeed confirmed by the data shown in Fig.~\ref{fig_diff}b.
The canonical scaling $D\propto \dot \gamma$ at low $\phi$ is characteristic for the purely hydrodynamic limit, where the external shear rate provides the only time scale of the problem \cite{leighton_shear-induced_1987,sierou_shear-induced_2004, brady_microstructure_1997}.
In contrast, in the jammed case, non-hydrodynamic effects, such as particle deformability \cite{loewenberg_collision_1997, lac_hydrodynamic_2007}, become relevant and give rise to anomalous scaling of the fluctuations as well as of the viscosity. 
Deviations seen in Fig.~\ref{fig_diff}b at large Ca might be related to the tumbling-to-tank-treading transition \cite{kruger_crossover_2013}.
We point out that, differently from \cite{mobius_relaxation_2010, sexton_bubble_2011}, but similar to \cite{besseling_three-dimensional_2007}, structural relaxation and macroscopic rheology are not connected by a simple Stokes-Einstein-type relation ($D\eta \sim \const$). 

The diffusivity grows approximately linearly with $\phi$ for large Ca and quadratically for low Ca (Fig.~\ref{fig_diff}c). 
These dependencies can be understood from the fact that rigid particles (small Ca) require three-particle collisions to break the time-reversal symmetry of Stokes flow, while two-particle collisions are sufficient when deformability is relevant (large Ca) \cite{leighton_shear-induced_1987,loewenberg_collision_1997, lac_hydrodynamic_2007}. 
Remarkably, for low capillary numbers, the diffusivity is a non-monotonic function of $\phi$ and reaches its maximum near the random-close-packing value of $\phi_c\simeq 0.66$, consistent with the idea of critical jamming point. 
The overall behavior is, by eq.~\eqref{eq_D_scal}, essentially a consequence of the $\phi$-dependencies of the velocity fluctuations $\uRMS$ and the empirical length scale $\lambda$, which both grow up to approximately $\phi_c\simeq 0.66$ and then slightly decrease.
For comparison, we remark that experiments on RBCs in quasi-2D pressure-driven flow \cite{higgins_statistical_2009} obtained $D/(\dot \gamma r^2)\simeq 0.05$ at $\Ca\simeq 0.008$ and $\phi \simeq 0.33$, which is of similar magnitude to our results.

\section{Stress fluctuations}

\begin{figure}[t]\centering
	(a)\includegraphics[width=0.71\linewidth]{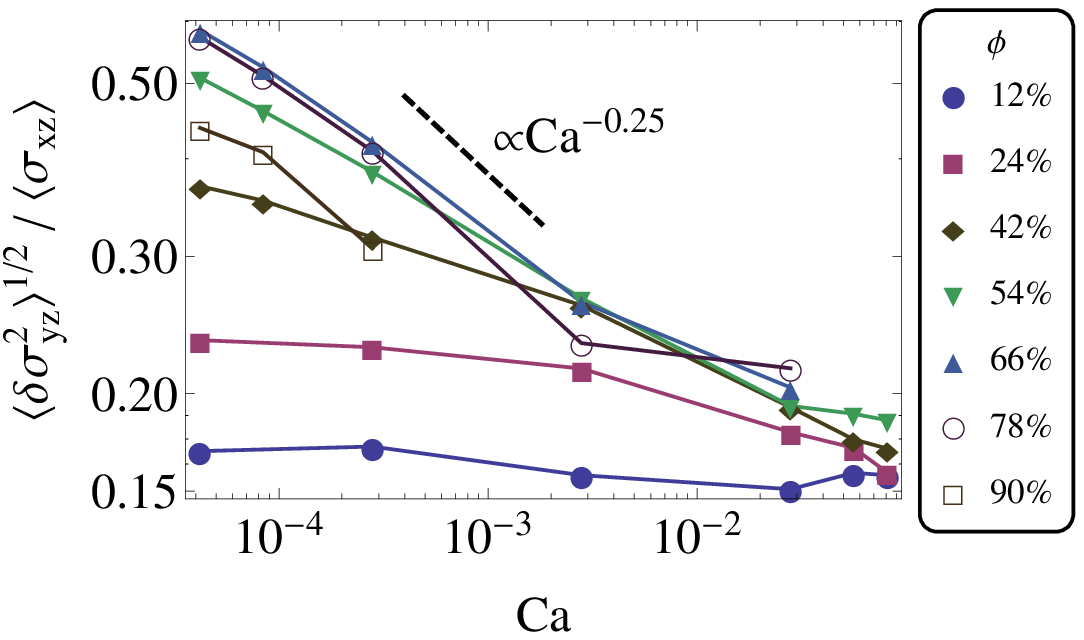}\quad
	(b)\includegraphics[width=0.62\linewidth]{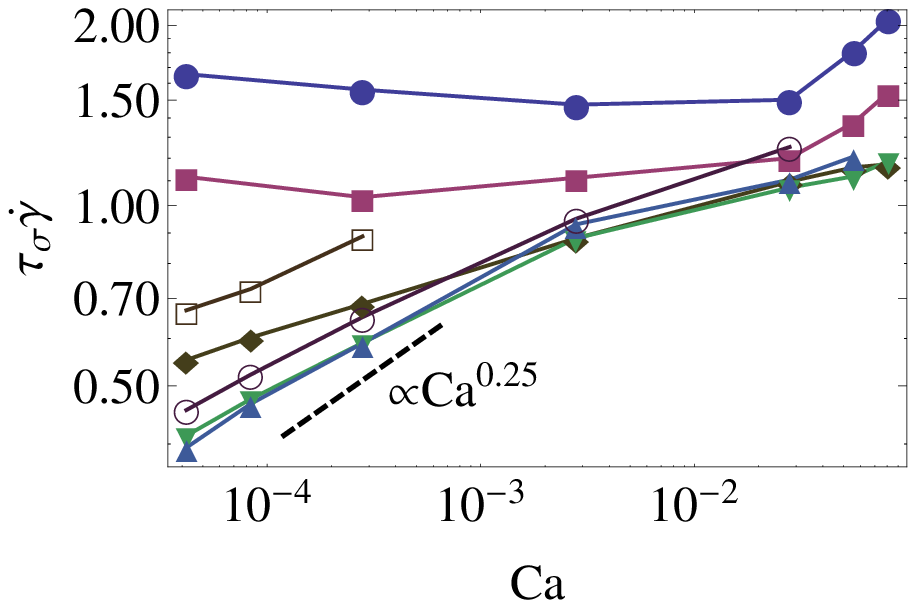}\quad
	\caption{(a) Root-mean-square of the particle stress fluctuations $\delta \sigma_{yz}$, normalized by the average particle stress in the suspension $\bra \sigma_{xz}\ket$. (b) Stress relaxation time obtained from stretched exponential fits to the stress autocorrelation function [legend in panel (a) applies].}
    \label{fig_stress_var}
\end{figure}

Owing to the deformability of the RBCs, mutual encounters generate elastic stresses in the cells which can affect their biochemical state \cite{hahn_mechanotransduction_2009}. 
Fig.~\ref{fig_stress_var}a shows the root-mean-square of the particle stress fluctuations $\delta \sigma_{yz}$, normalized by the average suspension stress $\bra\sigma_{xz}\ket=(\eta-\eta_0)\dot\gamma$ (cf.~\cite{gross_rheology_2014} for computational details). 
The observed scaling behavior can be rationalized by noting that the release of a stress fluctuation $\delta \sigma \simeq \bra\sigma^2\ket^\onehalf$ is associated with a driving force $F_\sigma \sim \delta\sigma\, d_\sigma^2$ ($d_\sigma$ being a characteristic length scale comparable to the cell diameter) that produces a displacement by a velocity $\delta u \simeq \uRMS$.  
Assuming the suspension to act as an effective medium of viscosity $\eta$, $F_\sigma$ will be counteracted by a drag force $F_d\sim d_\sigma \eta\, \delta u$. Balancing the two forces gives 
\beq d_\sigma \bra \sigma^2\ket^\onehalf \simeq \eta \uRMS\,.
\label{eq_stress_scal}
\eeq
As our simulations show that $d_\sigma$ varies only weakly over the studied parameter range, we find, using eq.~\eqref{eq_vel_scal}, that $\bra\sigma^2\ket^\onehalf/\bra\sigma\ket \propto \eta^{1/2}$, which is constant in the Newtonian regime and scales as $\Ca^{-q/2}$ in the shear-thinning regime, in good agreement with the results in Fig.~\ref{fig_stress_var}a. 
The bending of the curves at low Ca and large $\phi$ in Fig.~\ref{fig_stress_var}a might point to a different scaling behavior deep in the yield stress regime, where possibly $\bra\sigma^2\ket^\onehalf \sim \bra\sigma\ket$.
The coupling between stress and velocity relaxation furthermore suggests that the stress relaxation time $\tau_\sigma$ scales in a similar way as $\tau_v$ (see Fig.~\ref{fig_vel_relax}b). This is corroborated in Fig.~\ref{fig_stress_var}b, where $\tau_\sigma$ has been determined from fits of a stretched exponential decay to the stress autocorrelation function $\bra \delta\sigma_{xy}(t)\, \delta\sigma_{xy}(0)\ket$ at small strains.
A more detailed discussion of stress fluctuations will be presented elsewhere. 

\section{Summary and outlook}
We have shown that, besides the complex dynamics of the RBCs, their position and stress fluctuations generated in an athermal sheared setup can be understood based on a simple dissipation balance model, requiring only the viscosity as input. 
In agreement with our scaling arguments, extensive numerical simulations show that (reduced) velocity fluctuations $\uRMS/\dot \gamma$, stress fluctuations $\bra \sigma^2\ket^\onehalf/\sigma$, diffusivity $D/\dot \gamma$, velocity and stress relaxation rates $1/\tau_{v,\sigma}\dot \gamma$ all scale $\propto \eta^\onehalf$, with $\eta$ being the effective viscosity.
Remarkably, the detailed dynamics of the RBCs, such as the \TBTT\ \cite{kruger_crossover_2013}, do not seem to play a dominant role here.
This suggests the generality of our arguments, which can be straightforwardly applied to other types of suspensions or extended to include further dissipation mechanisms,  such as intra-cellular viscosity. 

Our results can be used to improve coarse-grained transport models \cite{phillips_constitutive_1992, kumar_mechanism_2012} describing particle migration and segregation effects.
We have demonstrated that deviations from the usually assumed canonical scaling behaviors are significant and must be taken into account when devising proper constitutive relations.
This should contribute to the better understanding of the role of particle fluctuations for nutrient transport, margination and occlusion phenomena in blood flow \cite{higgins_statistical_2009, kumar_mechanism_2012}.
We finally remark that, within the presently covered parameter region, the yield stress is small and clear effective power-laws could be identified.
For future work, it will be interesting to check our predictions further in the quasistatic regime, where finite-size effects are expected to become relevant \cite{lemaitre_rate-dependent_2009,varnik_correlations_2014} and the reduced fluctuation variances might cross over to another plateau.

\acknowledgments
\emph{Acknowledgments.---} We thank C.~Heussinger and V.~Koutsos for valuable discussions. This work is financially supported by the DFG-project Va205/5-2. We are also grateful for the computational time granted by the J\"ulich Supercomputing Centre (Project ESMI17).



\end{document}